\documentclass[12pt]{article}
\usepackage[dvips]{graphicx}

\def\PR#1#2#3{Phys. Rev. {\bf #1}, #2 (#3)}
\def\PRL#1#2#3{Phys. Rev. Lett. {\bf #1}, #2 (#3)}
\def\PL#1#2#3{Phys. Lett. {\bf #1}, #2 (#3)}

\def\NP#1#2#3{Nucl. Phys. {\bf #1}, #2 (#3)}

\def\PTP#1#2#3{Prog. Theor. Phys. {\bf #1}, #2 (#3)}
\def\EPJ#1#2#3{Eur.Phys.J. {\bf #1}, #2 (#3)}
\def\dfrac{\displaystyle\frac}
\def\mz{m_Z^{}}
\def\mr{m_R^{}}

\begin{document}
\title{The effects of Majorana phases in three-generation neutrinos}
\author{
{N. Haba$^{1,2}$}\thanks{E-mail address:haba@eken.phys.nagoya-u.ac.jp} 
{, Y. Matsui$^2$}\thanks{E-mail address:matsui@eken.phys.nagoya-u.ac.jp} 
{, and N. Okamura$^3$}\thanks{E-mail address:naotoshi.okamura@kek.jp} 
\\
\\
\\
{\small \it $^1$Faculty of Engineering, Mie University,}
{\small \it Tsu Mie 514-8507, Japan}\\
{\small \it $^2$Department of Physics, Nagoya University,}
{\small \it Nagoya, 464-8602, Japan}\\
{\small \it $^3$Theory Group, KEK, Tsukuba Ibaraki 305-0801, Japan}}
\date{}
\maketitle
\vspace{-11cm}
\begin{flushright}
hep-ph/0005075\\
DPNU-00-20\\
KEK-TH-692\\
\end{flushright}
\vspace{10.5cm}
\vspace{-2.5cm}
%
%
\begin{abstract}

Neutrino-oscillation solutions for
 the atmospheric neutrino anomaly and the solar neutrino deficit
 can determine the texture of the neutrino mass matrix
 according to three types of neutrino mass hierarchies as
 Type A: $m_1^{} \ll m_2^{} \ll m_3^{}$,
 Type B: $m_1^{} \sim m_2^{} \gg m_3^{}$ , and
 Type C: $m_1^{} \sim m_2^{} \sim m_3^{}$,
 where $m_i$ is the $i$-th generation neutrino absolute mass.
The relative sign assignments of neutrino masses in each type of
 mass hierarchies play the crucial roles for the stability against
 quantum corrections. 
Actually, two physical Majorana phases in the lepton
 flavor mixing matrix connect among
 the relative sign assignments of neutrino masses. 
Therefore, in this paper we analyze 
 the stability of mixing angles against quantum corrections 
 according to three types of neutrino mass hierarchies
 (Type A, B, C) and  two Majorana phases. 
Two phases play the crucial roles for the stability of 
the mixing angles against the quantum corrections.  

\end{abstract}


\newpage 

\section{Introduction}

Recent neutrino oscillation experiments suggest
 the strong evidences of tiny neutrino masses 
 and lepton flavor mixings\cite{solar4,Atm4,SK4,CHOOZ}.
Studies of the lepton flavor mixing matrix,
 which is so-called Maki-Nakagawa-Sakata(MNS) matrix\cite{MNS},
 will give us important cues of the physics
 beyond the standard model.
One of the most important studies is 
 the analysis of the quantum correction on 
 the MNS matrix\cite{up2now}--\cite{HMO1}.

In order to explain both the solar and the atmospheric neutrino
 problems, two mass squared differences are needed, 
 which implies
\begin{equation}
\Delta m_{\rm solar}^2 \equiv \left| m_2^2-m_1^2 \right| \,,
\mbox{\quad and \quad}
\Delta m_{\rm ATM}^2 \equiv \left| m_3^2-m_2^2 \right| \,,
\end{equation}
 where $m_i$ is the $i$-th ($i=1\sim 3$)
 generation neutrino mass ($m_i \geq 0$).
$\Delta m_{\rm solar}^2$ and $\Delta m_{\rm ATM}^2$ 
 stand for the mass-squared differences 
 of the solar neutrino \cite{solar4}
 and the atmospheric neutrino solutions\cite{Atm4,SK4}, 
 respectively.
Then there are the following three possible types of neutrino
mass hierarchies \cite{Altarelli};
\begin{eqnarray}
\mbox{Type A}
&~:~& 
m_1 \ll m_2 \ll m_3 \,,  \nonumber \\
\mbox{Type B}
&~:~& m_1 \sim m_2 \gg m_3\,, \\
\mbox{Type C}
&~:~& m_1 \sim m_2 \sim m_3\,, \nonumber 
\end{eqnarray}
where $m_i$ is the $i$-th generation neutrino absolute mass.
In Ref.~\cite{HO1}, 
 it has been studied whether
 the lepton-flavor mixing angles are stable or not 
 against quantum corrections
 for all three types of mass hierarchies 
 with all considerable relative sign assignments, which are shown below, 
 in the minimal supersymmetric standard model (MSSM)
 with an effective dimension-five operator which
 gives the Majorana masses of neutrinos.

\begin{enumerate}
\item  Type A: 
\begin{eqnarray}
 {\rm case (a1): \;\;} m_{\nu}^{\rm a1} &=& diag.(0, m_2^{},m_3^{})\,, 
\label{eqn:a1}\\
 {\rm case (a2): \;\;}m_{\nu}^{\rm a2} &=& diag.(0,-m_2^{},m_3^{})\,.
\end{eqnarray}
\[
\left(\,\,
m_1  =  0, \;\;\;\;\;\,
m_2  =  \sqrt{\Delta m_{\rm solar}^2}, \;\;\;\;\;\,
m_3  =  \sqrt{\Delta m_{\rm solar}^2 + \Delta m_{\rm ATM}^2}
\,\,
\right) \, 
\]
\item Type B:  
\begin{eqnarray}
 {\rm case (b1): \;\;} m_{\nu}^{\rm b1} &=& diag.(m_1^{}, m_2^{},0)\,, \\
 {\rm case (b2): \;\;}m_{\nu}^{\rm b2} &=& diag.(m_1^{}, -m_2^{},0)\,.
\end{eqnarray}
\[
\left(\,\,
m_1  = \sqrt{\Delta m_{\rm ATM}^2} , \;\;\;\;\;\,  
m_2  =  \sqrt{\Delta m_{\rm solar}^2 + \Delta m_{\rm ATM}^2}, \;\;\;\;\;\,
m_3  =  0
\,\,
\right)
\]
\item Type C:  
\begin{eqnarray}
\mbox{case (c1):\quad} m_{\nu}^{c1} &=&
 diag.(-m_1^{}, m_2^{},m_3^{})\,,  \\
\mbox{case (c2):\quad} m_{\nu}^{c2} &=&
 diag.( m_1^{},-m_2^{},m_3^{})\,,  \\
\mbox{case (c3):\quad} m_{\nu}^{c3} &=&
 diag.(-m_1^{},-m_2^{},m_3^{})\,,  \\
\mbox{case (c4):\quad} m_{\nu}^{c4} &=&
 diag.( m_1^{}, m_2^{},m_3^{})\,. \label{eqn:c4}
\end{eqnarray}
\[
\left(
\,\,
m_1^{} = m_0^{} \,, \mbox{\quad}
m_2^{} = \sqrt{m_0^2+\Delta m_{\rm solar}^2}\,, \mbox{\quad}
m_3^{} = \sqrt{m_0^2+\Delta m_{\rm solar}^2+\Delta m_{\rm ATM}^2}
\,\,
\right)
\]
\end{enumerate}
In Ref.\cite{HO1}, it has been found that 
 the above relative sign assignments of neutrino masses in each type
 play the crucial roles for the stability of the mixing angles 
 against quantum corrections. 
Actually, two physical Majorana phases in the lepton
 flavor mixing matrix connect among
 the above relative sign assignments of neutrino masses. 
Therefore, in this paper we analyze 
 the stability of mixing angles against 
 quantum corrections 
 according to three types 
 of neutrino mass hierarchies (Type A, B, C) and 
 two Majorana phases. 
Two phases play the crucial roles for the stability of 
 the mixing angles against the quantum corrections.
In Refs.~\cite{HMOS2,HMO1}, 
 it has been already analyzed that 
 the effect of a Majorana phase plays an important role
 for the stability against the quantum corrections
 in the two-generation neutrinos. 

\section{Quantum corrections to neutrino mass matrix}

In the MSSM with
 the effective dimension-five operator
 which gives Majorana masses of neutrinos, 
 the superpotential of the lepton-Higgs interactions is
 given by 
\begin{equation}
{\cal W} =
 y^{\rm e}_{ij} (H_{d} L_i) {E_{j}}
 - \dfrac{1}{2} \kappa^{}_{ij} (H_{u} L_i)(H_{u} L_j) \,.
\label{superpot}
\end{equation}
Here the indices $i,j$ $(=1 \sim 3)$ stand for the generation number.
 $L_i$ and $E_{i}$ are chiral super-fields of $i$-th generation 
 lepton doublet and right-handed charged-lepton, respectively.
 $H_{u}$ ($H_{d}$) is the Higgs doublet
 which gives Dirac masses to the up- (down-) type fermions. 
The neutrino mass matrix of the three generations,
$\kappa$ is diagonalized as
\begin{equation}
 U^{T} \kappa \ U = D_{\kappa}\,,
\label{d-kappa}
\end{equation}
where $D_{\kappa}$ is given by 
\begin{equation}
 D_{\kappa} =
\left(
\begin{array}{ccc}
 m_1 & 0 & 0 \\
 0 & m_2 & 0 \\
 0 &   0 & m_3
\end{array}
\right)\,,
\label{mi2}
\end{equation}
with $m_i \geq 0$. 
The unitary matrix $U$ 
 is defined as
\begin{equation}
\label{MNS}
U = 
\left(
\begin{array}{ccc}
 U_{e1} & U_{e2} & U_{e3} \\
 U_{\mu 1} & U_{\mu 2} & U_{\mu 3} \\
 U_{\tau 1} & U_{\tau 2} & U_{\tau 3}
\end{array}
\right)
\left(
\begin{array}{ccc}
e^{i \phi_1}   & 0   & 0         \\
0   & e^{i \phi_2}   & 0         \\
0   & 0              &  1 
\end{array}
\right) \,,
\end{equation}
where $\phi_{1,2}$ denote the physical Majorana
 phases of the lepton sector. 
In the diagonal base of charged lepton masses, 
 $U$ is just the MNS matrix.
We can easily show that 
 one Majorana phase connects 
 between cases of (a1) and (a2), (b1) and (b2), and  
 two Majorana phases connect 
 among cases of (c1)$\sim$(c4). 
Thus, the stabilities of mixing angles against 
 quantum corrections 
 are completely determined by three types 
 of neutrino mass hierarchies (Type A, B, C) and 
 two Majorana phases $\phi_{1,2}$ in stead of the classifications of 
 Eqs. (\ref{eqn:a1}) $\sim$ (\ref{eqn:c4}).

We will analyze whether the lepton flavor mixing angles 
 are changed or not by the quantum corrections
 by fitting the low energy data. 
We determine the MNS matrix at $\mz$ scale as 
\begin{equation}
\label{UUU}
U = 
\left(
\begin{array}{ccc}
 \cos \theta _{12}  & \sin \theta _{12} & 0 \\
  - \dfrac{\sin \theta _{12}}{\sqrt{2}} & 
    \dfrac{\cos \theta _{12}}{\sqrt{2}} & \dfrac{1}{\sqrt{2}} \\
    \dfrac{\sin \theta _{12}}{\sqrt{2}} & 
  - \dfrac{\cos \theta _{12}}{\sqrt{2}} & \dfrac{1}{\sqrt{2}} 
\end{array}
\right)
\left(
\begin{array}{ccc}
e^{i \phi_1}   & 0   & 0        
\bigskip \\
0   & e^{i \phi_2}   & 0       
\bigskip \\
0   & 0              &  1 
\end{array}
\right) \,,
\end{equation}
where we input $\sin \theta _{23} = 1 / \sqrt{2}$ and $\sin \theta _{13} = 0$
which values are suitable for 
 the atmospheric neutrino  experiments \cite{Atm4,SK4} and
 for the CHOOZ experiment \cite{CHOOZ}, respectively.
The mixing angle $\theta _{12}$ depends on the  
solar neutrino solutions of
the large angle MSW solution (MSW-L), the small angle MSW solution (MSW-S) 
and vacuum oscillation solution (VO), which are given by
\begin{eqnarray}
\sin \theta _{12}= 
\left\{
\begin{array}{lll}
0.0042   & ( \theta = 0.0042 ) &  \mbox{(MSW-S), } \\
{1 \over \sqrt{2}} & ( \theta = {\pi \over 4} ) &  \mbox{(MSW-L),}  \\
{1 \over \sqrt{2}} & ( \theta = {\pi \over 4} ) &  
         \mbox{(VO).}  \\
\end{array}
\right. 
\end{eqnarray}

We also use the following values of mass-squared differences 
 in the numerical analyses. 
\begin{eqnarray}
\Delta m_{\rm solar}^2 &\simeq&
\left\{
\begin{array}{cl}
0.8  \times 10^{-5} &  \mbox{eV$^2$~~(MSW-S),}  \\
1.8  \times 10^{-5} &  \mbox{eV$^2$~~(MSW-L),}  \\
0.85 \times 10^{-10}&  \mbox{eV$^2$~~(VO), } \\
\end{array}
\right. \\
\Delta m_{\rm ATM}^2 & \simeq &3.7  \times 10^{-3}  \mbox{eV$^2$} \, .
\end{eqnarray}

The quantum corrections 
 change the neutrino mass matrix, and 
 it is given by\footnote{%
Hereafter, we denote the mixing angles and the other 
physical parameters at the $\mr$ scale
 are written with $\hat{\hspace{1ex}}$ mark.
}\cite{EL1,HMOS1}
\begin{eqnarray}
\label{k'}
\hat{\kappa}\left(\mr\right)
&=& \dfrac{\hat{\kappa}\left(\mr\right)_{33}}
          {\kappa\left(\mz\right)_{33}}
\left(
\begin{array}{ccc}
 1- \epsilon    &  0           & 0   \\
 0              & 1- \epsilon  & 0   \\
 0              &  0           & 1
\end{array}
\right)
\kappa \left(\mz\right)
\left(
\begin{array}{ccc}
 1- \epsilon    &  0           & 0   \\
 0              & 1- \epsilon  & 0   \\
 0              &  0           & 1
\end{array}
\right)\,, 
\end{eqnarray}
at the high energy scale $\mr$, where $\epsilon$ can be estimated as
\begin{eqnarray}
\label{def_eta}
\epsilon & \simeq & 1 -
{\exp
\left(-\dfrac{1}{16\pi^2}
\int^{\ln\left(\mr\right)}_{\ln\left(\mz\right)}
y_\tau^2 dt
\right)} ,  \nonumber \\
& \simeq &
\dfrac{1}{8\pi^2}
\dfrac{m_\tau^2}{v^2}\left(1+\tan^2\beta\right)
\ln \left(\dfrac{\mr}{\mz}\right)\,.
\end{eqnarray}
where $y_\tau$ is the Yukawa coupling of $\tau$, 
$v^2 \equiv v_u^2 + v_d^2$ and $\tan \beta \equiv v_u / v_d$
($v_u$ and $v_d$ are vacuum expectation values of Higgs bosons,
$H_u$ and $H_d$, respectively).
We neglect the Yukawa couplings
 of $e$ and $\mu$ in Eqs.(\ref{k'}) and (\ref{def_eta}), 
 since those contributions to the renormalization group equations
 are negligibly small comparing to that of $\tau$ \cite{HO1}.
The magnitude of $\epsilon$ can be determined by 
 the value of $\tan \beta$
  and the scale of $\mr$. 
The unitary matrix $\hat{U}$ which diagonalizes
 $\hat{\kappa}$ shows us
 whether the lepton flavor mixing angles 
 are stable against quantum corrections or not.

\section{Type A $(m_1 \ll m_2 \ll m_3)$}

In both (a1) and (a2) cases, all mixing angles are stable 
 against quantum corrections in each sign assignment \cite{HO1}. 
This is understood from the analogy of two-generation
 analysis,  which shows 
 the mixing angle of 
 $2 \times 2$ mass matrix is not changed drastically by 
 the quantum corrections 
 when there is the large mass hierarchy between
 two neutrinos \cite{HO1}. 
This situation is not changed when we consider the Majorana phase
 contribution as shown in two-generation neutrinos \cite{HMOS2}. 
Cases (a1) and (a2) are connected with each other 
 by the Majorana phase of $\phi_2$. 
Where $\phi_1$ is rotated out, since $m_1=0$. 
The case of $\phi_2=0$ corresponds to (a1), 
 while the case of $\phi_2 = \pi /2$ corresponds to (a2). 
Since Type A has the large mass hierarchies, 
 all mixing angles are supposed to be stable against quantum corrections
 independently of the value of the Majorana phase $\phi_2$. 
This is really confirmed by numerical analyses as shown in Table
\ref{tbl:typeA},
where we use $\mr=10^{13}$ GeV and $\tan \beta = 60$.

\begin{table}[htb]
\begin{center}
\begin{tabular}{|c|c|c|c|}
\cline{2-4}
\multicolumn{1}{c|}{ } & MSW-S & MSW-L & VO \\
\hline
$\sin^2 2 \hat{\theta}_{12}$ & $0.005$ & $0.998$ & $0.998$ \\
\hline
$\sin^2 2 \hat{\theta}_{23}$ & $0.985 \sim 0.99$ & $0.985 \sim 0.99$ & $0.99$ \\
\hline
$\sin^2 2 \hat{\theta}_{13}$ & $10^{-7}$ & $10^{-4}$ & $10^{-10}$ \\
\hline
\end{tabular}
\end{center}
\caption{Stabilities of the mixing angles
with the Type A mass hierarchy according to the change of $\phi_2$
 from $0$ to $\pi$ in the case of $\mr=10^{13}$ GeV and $\tan \beta = 60$.}
\label{tbl:typeA}
\end{table}

\section{Type B $( m_1 \sim m_2 \gg m_3)$}

In Type B mass hierarchy, all mixing angles 
 except for $\sin \theta_{12}$ of (b2) are stable 
 against quantum corrections \cite{HO1}. 
The analogy of two-generation neutrinos
 analysis shows that mixing angles of $\sin \theta_{13}$ and 
 $\sin \theta_{23}$ are stable against quantum corrections, 
 since there are large mass hierarchies between 
 the first and the third generations, and 
 between the second and third generations. 
This is the same situation as that of Type A. 
This situation is not changed by including 
 the Majorana phase contributions of $\phi_{1,2}$
 as shown in Table \ref{tbl:typeB}, which shows the 
 results of the numerical analyses in the case of
 $\mr=10^{13}$ GeV and $\tan \beta=60$.
\begin{table}[htb]
\begin{center}
\begin{tabular}{|c|c|c|c|}
\cline{2-4}
\multicolumn{1}{c|}{} & MSW-S & MSW-L & VO \\
\hline
$\sin^22\hat{\theta}_{12}$ &
  \multicolumn{1}{|c}{} & \multicolumn{1}{c}{See Figures \ref{fig:typeB1}, 
  \ref{fig:typeB2}} & \multicolumn{1}{c|}{} \\
\hline
$\sin^22\hat{\theta}_{23}$ & 0.99 & 0.99 & 0.99 \\
\hline
$\sin^22\hat{\theta}_{13}$ & 0 & 0 & 0 \\
\hline
\end{tabular}
\end{center}
\caption{Stabilities of the mixing angles
with the Type B mass hierarchy according to the change of $\phi$ from
 $0$ to $\pi$. In this analysis we use the $\mr=10^{13}$ GeV and
$\tan \beta=60$.}
\label{tbl:typeB}
\end{table}
On the other hand, the mixing angle of $\theta_{12}$ can receive significant
 quantum corrections dependently on the relative sign assignment 
 of $m_2$ as shown in Ref \cite{HO1}. 
The mixing angle of $\sin \theta_{12}$ of (b1) receives 
 the quantum correction 
 while that of (b2) does not. 
Now we understand that two cases of (b1) and (b2)  
 are connected with each other by the phase of 
 $\phi \equiv \phi_1 - \phi_2$, which 
 is the only physical phase, since $m_3=0$. 
The case of $\phi=0$ corresponds to (b1), while 
 the case of $\phi = \pi /2$ corresponds to (b2). 
The phase $\phi$ is the parameter which 
 determines whether the mixing angle $\theta_{12}$ is stable 
 against quantum corrections or not.

Now let us show the analytic estimations for the stabilities of the mixing angles
in  Type B mass hierarchy.
 The neutrino mass matrix of Type B which is diagonalized  
 is given by 
\begin{equation}
 D_{\kappa}^{(B)} = m_1 
\left(
\begin{array}{ccc}
 1 & 0 & 0 \\
 0 & 1+ \xi_b & 0 \\
 0 &   0 & 0
\end{array}
\right)\,,
\label{mi}
\end{equation}
where
\begin{equation}
 \xi_b \equiv {m_2 - m_1 \over m_1} \simeq  {1 \over 2} 
 {\Delta m_{\rm solar}^2 \over \Delta m_{\rm ATM}^2}\, .  
\end{equation}
We can determine the mass matrix of $\kappa ^{(B)}$ 
 by using Eqns.(\ref{d-kappa}) and (\ref{UUU}). 
Then Eq.(\ref{k'}) gives the mass matrix of $\hat{\kappa} ^{(B)}$
at the high energy scale $\mr$.

The MNS matrix $\hat{U} ^{(B)}$ which 
 diagonalizes $\hat{\kappa} ^{(B)}$ is given by
\begin{eqnarray}
 \hat{U} ^{(B)} &=& 
\left(
\begin{array}{ccc}
 1 &      0 &        0 \\
 0 & (1- \epsilon)/ \sqrt{1+(1- \epsilon )^2}  & 1/ \sqrt{1+(1- \epsilon )^2}  \\
 0 &  -1/ \sqrt{1+(1- \epsilon )^2}  & (1- \epsilon)/ \sqrt{1+(1- \epsilon )^2}
\end{array}
\right)
\nonumber \\
& & \;\;\;\;\;\;\;\;\;\;\;\;\;\;\;\;\;\;\;\;\;\; \times
\left(
\begin{array}{ccc}
 \cos \hat{\theta}_{12}   & \sin \hat{\theta}_{12}    &        0 \\
 - \sin \hat{\theta}_{12} & \cos \hat{\theta}_{12}    &        0 \\
       0             &       0             &        1  
\end{array}
\right)
\left(
\begin{array}{ccc}
e^{i \hat{\phi}_1}   & 0   & 0         \\
0   & e^{i \hat{\phi}_2}   & 0         \\
0   & 0              &  1 
\end{array}
\right) \,,
\end{eqnarray}
which means that the mixing angle between the first and the third generations,
 which is zero, is unchanged by quantum corrections.
The mixing angle of $\hat{\theta} _{23}$ is given by
\begin{equation}
  \sin^2 2 \hat{\theta} _{23} = 
\left(\dfrac{2 (1 - \epsilon)}{1 + (1-\epsilon)^2}\right)^2 \;\;,
\label{typeBs23}
\end{equation}
which indicates that the large mixing between the 
 second and the third generations 
 is stable with respect to quantum corrections.
 By using Eq.(\ref{typeBs23}), we can estimate 
 that $\sin^2 2 \hat{\theta}_{23} \simeq 0.99 $
 in the case of $\mr=10^{13}$ GeV and $\tan \beta = 60$,  
 which is consistent with the numerical analysis in Table \ref{tbl:typeB}.
Therefore the mixing between the first and the third generations 
and the mixing between the second and the third generations are stable 
with respect to
quantum corrections as shown in Table \ref{tbl:typeB}

How about the mixing between the first and the second generations?
\begin{figure}[tb]
\begin{center}
 {\scalebox{0.8}{\includegraphics{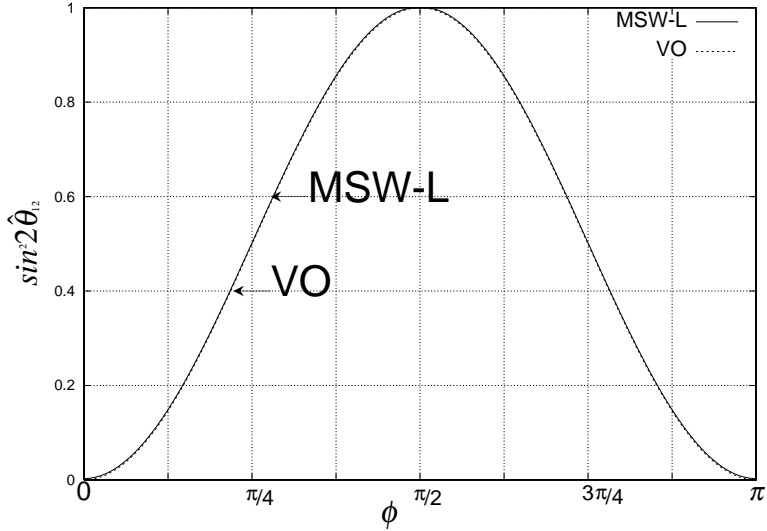}} } 
\end{center}
 \vspace{-1.5em}
 \caption{
Majorana phase dependence of $\sin^2 2 \hat{\theta}_{12}$
for the MSW-L and the VO solutions  in Type B mass hierarchy
in the case of $\mr = 10^{13}$ GeV and $\tan \beta =60$.
}
\label{fig:typeB1}
\end{figure}

\noindent
For the MSW-L and the VO solutions, 
 where $\sin \theta _{12} = \cos \theta _{12} = 1 / \sqrt{2}$ 
 at $\mz$ scale,  
 the mixing angle of $\tan \hat{\theta}_{12}$
 is given by 
\begin{eqnarray}
\tan 2 \hat{\theta}_{12} 
& \simeq & 
(1- \epsilon) 
\sqrt{1- \epsilon} \;\;
{\sqrt{4 \xi_b^2 + \epsilon^2 \sin^2 2 \phi} \over \epsilon (1+ \cos 2 \phi)} \;,
\label{tan12}
\end{eqnarray}
where we use the approximation which neglects 
 the higher order corrections of $\epsilon^2$, 
 $\epsilon \xi_b$, and $\xi_b^2$. 
When $\phi = \pi / 2$, the mixing angle $\hat{\theta}_{12}$ becomes 
\begin{equation}
\tan 2 \hat{\theta}_{12} = \infty , 
\end{equation}
which means the maximal mixing is stable 
 against quantum corrections. 
On the other hand, 
 when $\phi = 0$, 
\begin{equation}
\tan 2 \hat{\theta}_{12} \simeq {\xi_b \over \epsilon}, 
\end{equation}
which shows that the mixing angle of $\hat{\theta}_{12}$ strongly depends 
 on the magnitude of $\epsilon$. 
The large mixing is spoiled when $\xi_b \leq \epsilon$, which corresponds the region of
 $\tan \beta \geq 10$ for the MSW-L solution, and
any value of $\tan \beta$ for the VO solution when we take $\mr = 10^{13}$ GeV. 
In Fig.\ref{fig:typeB1},  we show the change of 
 $\sin^2 2 \hat{\theta} _{12}$ due to 
the continuous change of Majorana phase $\phi$ 
in the case of $\tan \beta = 60$ and 
$\mr = 10 ^{13}$ GeV.
As the Majorana phase $\phi$ changes  from 
$0$ to $\pi/2$, the value of $\sin^2 2 \hat{\theta}_{12}$
 changes from $0$ to $1$.
The large deviation from $1$ of $\sin^2 2 \hat{\theta}_{12}$ means 
  that the mixing angle $\theta_{12}$
  is unstable with respect to the quantum corrections.
Figure \ref{fig:typeB1} shows that the mixing angle $\theta_{12}$ 
changes from being unstable to being stable as 
the change $\phi$ from $0$ to $\pi$.
The lines of the MSW-L and the VO solutions are almost overlapping
in Fig.\ref{fig:typeB1},
since the discrepancy of $\xi _b$ 's for the two solutions is
negligible compared with the quantum correction, $\epsilon=0.1$, 
when  $\tan \beta = 60$ and  $\mr = 10 ^{13}$ GeV.

\begin{figure}[tb]
\begin{center}
 {\scalebox{0.8}{\includegraphics{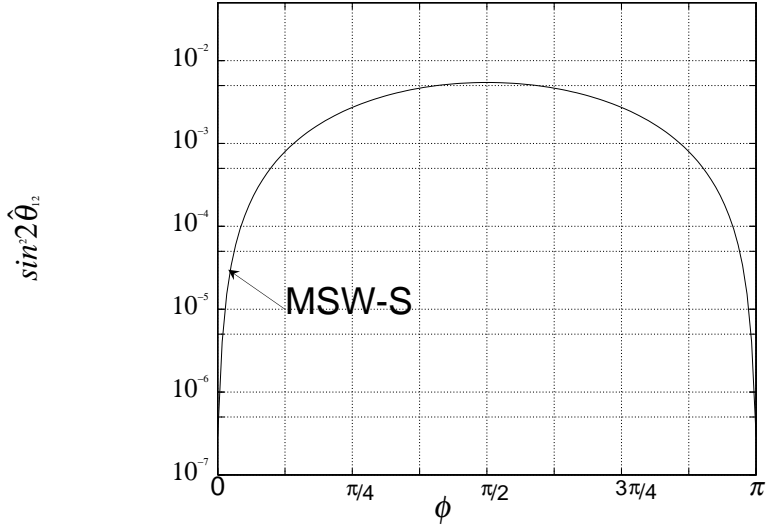}} } 
\end{center}
 \vspace{-1.5em}
 \caption{
  Majorana phase dependence of the $\sin^2 2 \hat{\theta}_{12}$
  for the MSW-S solution in Type B mass hierarchy in the
  case of  $\tan \beta = 60$ and  $\mr = 10 ^{13}$ GeV.
}
\label{fig:typeB2}
\end{figure}

As for the MSW-S solution, $\phi=0$ induces 
\begin{equation}
\label{34}
\tan 2 \hat{\theta}_{12} \simeq 
 \tan 2 \theta_{12} \left( 1 +  {1 \over \cos 2 \theta_{12}}
{\epsilon \over \xi_b} \right)^{-1},
\end{equation}
while $\phi= \pi / 2$ induces 
\begin{equation}
\label{35}
\tan 2 \hat{\theta}_{12} \simeq 
 \tan 2 \theta_{12} \;.
\end{equation}
Equations (\ref{34}) and (\ref{35}) show that 
 the mixing angle of $\theta_{12}$ is not changed 
 in the region of $\tan \beta \leq 10$ when $\phi =0$, 
 while it is not changed independently  of $\tan \beta$ when 
 $\phi = \pi / 2$.
Above conclusions are the same as those of Ref.\cite{HO1}.
In Fig. \ref{fig:typeB2}, we show the value of $\sin^2 2 \hat{\theta}_{12}$ 
 at $\mr = 10^{13}$ GeV scale in the case of $\tan \beta = 60$ 
 according to  the continuous change of $\phi$ from $0$ to $\pi$. 
Figure \ref{fig:typeB2} shows that the mixing angle $\theta_{12}$ 
 changes from being unstable to being stable corresponding 
 to the change of $\phi$ from $0$ to $\pi$.

\section{Type C $(m_1 \sim m_2 \sim m_3)$}

In Type C mass hierarchy, 
it has been shown in Ref.\cite{HO1} that 
 the MNS matrix approaches the definite unitary matrix according to the
 relative sign assignments of the neutrino mass eigenvalues,
 as the effects of quantum corrections become
 large enough to neglect the mass-squared differences of neutrinos.
Independent parameters of the MNS matrix at the $\mr$ scale
 approach the following fixed values in the large limit of 
 quantum corrections:

\noindent
\underline{case (c1)}: $diag.(-m_1^{}, m_2^{},m_3^{})$ 
\begin{equation}
U_{e2}=\dfrac{\sin\theta_{12}}{\sqrt{1+\cos^2\theta_{12}}}\,,
\mbox{\quad}
U_{e3}=-\dfrac{1}{2}\dfrac{\sin 2\theta_{12}}{\sqrt{1+\cos^2\theta_{12}}}\,,
\mbox{\quad}
U_{\mu 3}=\dfrac{1}{\sqrt{2}}\dfrac{\sin^2\theta_{12}}{\sqrt{1+\cos^2\theta_{12}}}\,. 
\label{eqn:Cc1}
\end{equation}

\noindent
\underline{case (c2)}: $diag.( m_1^{},-m_2^{},m_3^{})$ 
\begin{equation}
U_{e2}=\sin \theta_{12}\,,
\mbox{\quad}
U_{e3}=\dfrac{1}{2}\dfrac{\sin 2 \theta_{12}}{\sqrt{1+\sin^2\theta_{12}}}\,,
\mbox{\quad}
U_{\mu 3}=\dfrac{1}{\sqrt{2}}\dfrac{\cos^2\theta_{12}}{\sqrt{1+\sin^2\theta_{12}}}\,.
\end{equation}

\noindent
\underline{case (c3)}: $diag.(-m_1^{},-m_2^{},m_3^{})$ 
\begin{equation}
U_{e2}=0\,,
\mbox{\quad}
U_{e3}=0\,,
\mbox{\quad}
U_{\mu 3}=\dfrac{1}{\sqrt{2}}\,.
\end{equation}

\noindent
\underline{case (c4)}: $diag.( m_1^{}, m_2^{},m_3^{})$ 
\begin{equation}
U_{e2}=0\,,
\mbox{\quad}
U_{e3}=0\,,
\mbox{\quad}
U_{\mu 3}=0\,. \label{eqn:Cc4}
\end{equation}
We can easily obtain the values of the mixing angles by
 using relations of \cite{HagiwaraOkamura},
\begin{eqnarray}
  \sin^2 2\theta_{12} &=& 4 \dfrac{U_{e2}^2}{1-|U_{e3}|^2}
                             \left(1-\dfrac{U_{e2}^2}{1-|U_{e3}|^2}\right)\,,
  \label{sin2_12}\\
  \sin^2 2\theta_{13} &=& 4 |U_{e3}|^2 \left(1-|U_{e3}|^2\right)\,, 
  \label{sin2_13} \\
  \sin^2 2\theta_{23} &=& 4 \dfrac{U_{\mu 3}^2}{1-|U_{e3}|^2}
                      \left(1-\dfrac{U_{\mu 3}^2}{1-|U_{e3}|^2}\right)\,.
  \label{sin2_23}
\end{eqnarray}
As shown above,  the cases of (c1)$\sim$(c4) are 
 connected by Majorana phases 
 of $\phi_1$ and $\phi_2$. 

\begin{figure}[tb]
\begin{center}
  {\scalebox{0.8}{\includegraphics{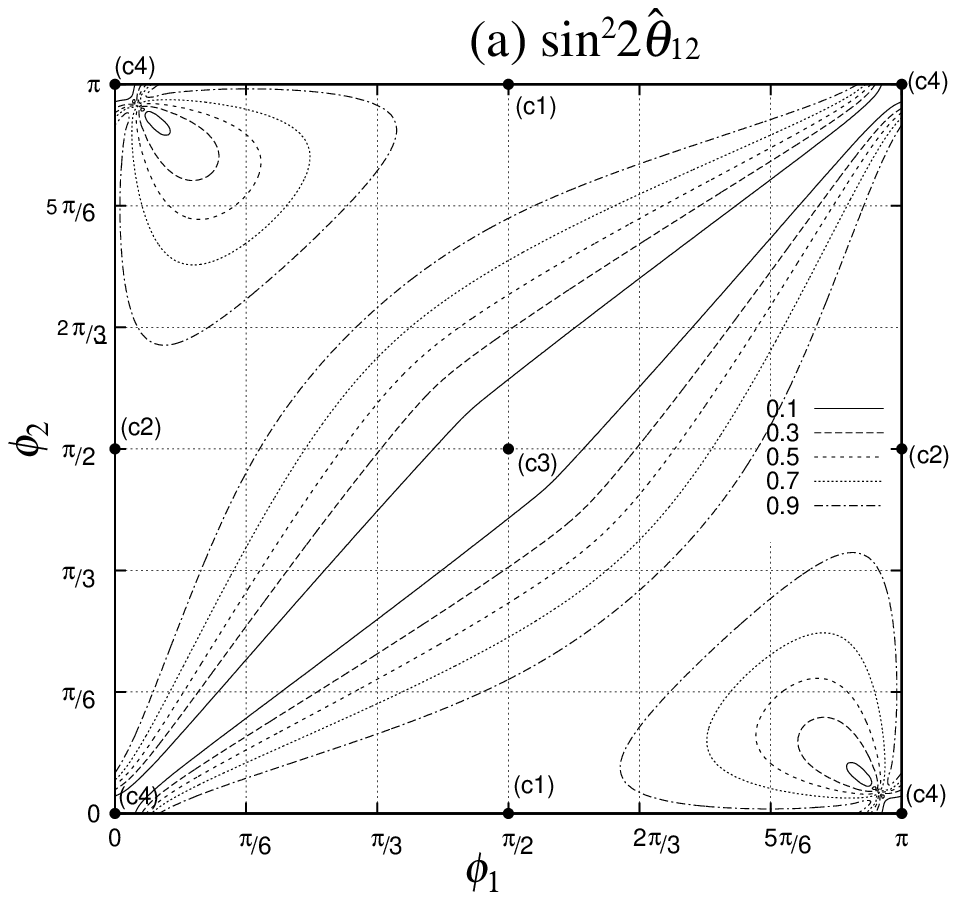}} } 
  {\scalebox{0.8}{\includegraphics{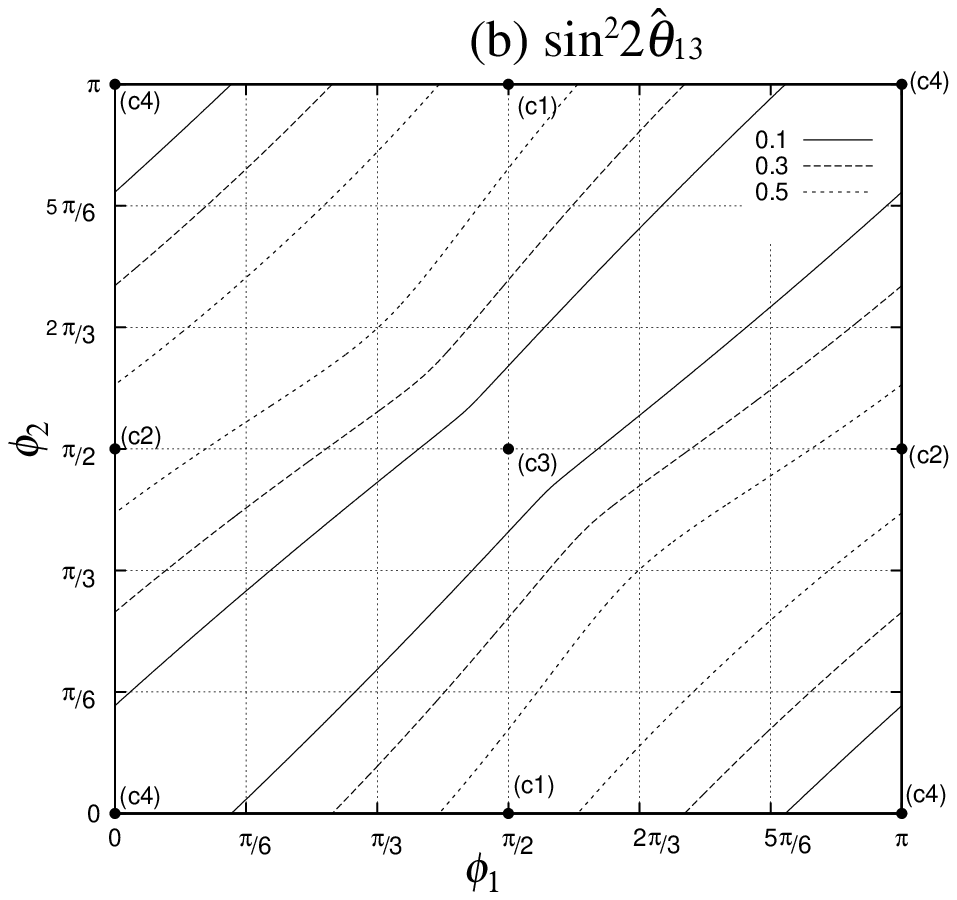}} } 
  {\scalebox{0.8}{\includegraphics{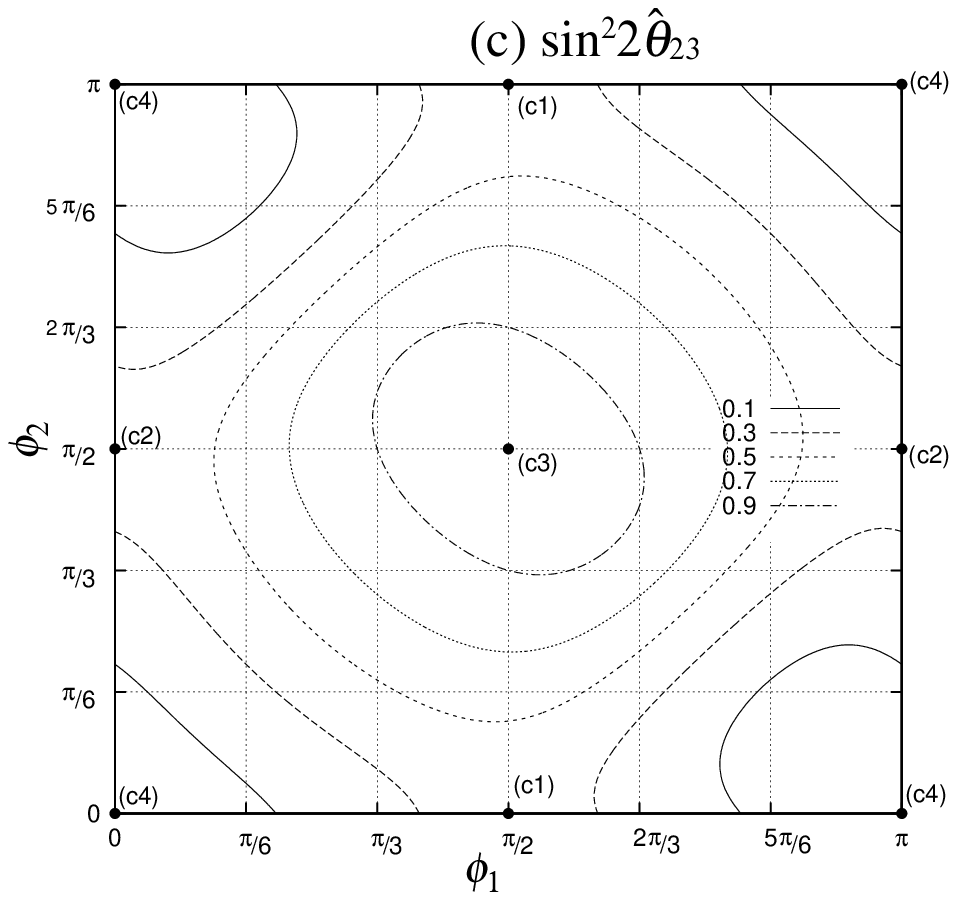}} } 
\end{center}
 \vspace{-1.5em}
 \caption{The contour plots of (a): $\sin^2 2 \hat{\theta} _{12}$,
          (b): $\sin^2 2 \hat{\theta} _{13}$ and (c): $\sin^2 2 
          \hat{\theta} _{23}$, 
          at $\mr = 10^{13}$ GeV in the case of
          the MSW-L and the VO solutions with $\tan \beta = 60$. }
 \label{fig:C-MSWL} 
\end{figure}
Figure \ref{fig:C-MSWL} shows that the values of mixing angles
 at the high energy scale $\mr = 10^{13}$ GeV 
 for the MSW-L and the VO solutions according to continuous
 changes of Majorana phases $\phi _1$ and $\phi _2$
 in the case of $\tan \beta = 60$.
Under the conditions that the effects of quantum corrections are
 large enough to neglect the mass-squared differences of neutrinos,
 the results of the MSW-L solution are the same as those of the VO
 solution \cite{HO1}.
Table \ref{tbl:C-MSWL} shows the fixed values of the 
 mixing angles for the MSW-L and the VO solutions in
 the large limit of quantum corrections which are
 obtain form Eqns.(\ref{eqn:Cc1}) $\sim$ (\ref{eqn:Cc4}) by
 using Eqns.(\ref{sin2_12}) $\sim$ (\ref{sin2_23}).
%
%
\begin{table}[htb]
\begin{center}
\begin{tabular}{|c||c|c|c|c|}
\cline{2-5}
\multicolumn{1}{c|}{} & \multicolumn{1}{c|}{(c1)} & (c2) & (c3) & (c4) \\
\hline
$\sin^2 2 \hat{\theta}_{12}$ & $0.96$ & $0.96$ & $0.0$ & $0.0$ \\
\hline
$\sin^2 2 \hat{\theta}_{13}$ & $0.56$ & $0.56$ & $0.0$ & $0.0$ \\
\hline
$\sin^2 2 \hat{\theta}_{23}$ & $0.36$ & $0.36$ & $1.0$ & $0.0$ \\
\hline
\end{tabular}
\end{center}
\caption{The fixed values of the mixing angles for the MSW-L and
         the VO solutions in the large limit of quantum corrections given by 
         Eqns.(\ref{eqn:Cc1}) $\sim$ (\ref{sin2_23}).
}
\label{tbl:C-MSWL}
\end{table}
The deviations from the values at $\mz$ scale, 
 $\sin^2 2 \theta _{12} = 1$, $\sin^2 2 \theta _{13} = 0$ and
 $\sin^2 2 \theta _{23} = 1$, indicate that mixing angles
 receive significant quantum corrections.
For $\sin^2 2 \hat{\theta}_{12}$, Table \ref{tbl:C-MSWL} shows
 that the cases of (c1) and (c2) conserve the maximal mixing,
 while the cases of (c3) and (c4) do not, 
 in the large limit of quantum corrections.
From Eq.(\ref{UUU}), we can show that the change of $\phi _1$ form 
 $0$ to $\pi/2$ with the relation $\left| \phi _2 - \phi _1 \right| = 0$
 ($\left| \phi _2 - \phi _1 \right| = \pi/2$) corresponds to
 the change of (c4) to (c3) ((c2) to (c1)).
Figure \ref{fig:C-MSWL}(a) shows that the unstable region of 
 $\sin^2 2 \hat{\theta}_{12} \mathop{}^{<}_{\sim} 0.1$ 
  exists around the line of $\left| \phi _2 - \phi _1 \right| = 0$,
  and the stable region of $\sin^2 2 \hat{\theta}_{12} \sim 1.0$
  exists around the line of $\left| \phi _2 - \phi _1 \right| = \pi/2$.
Since the cases of (c1) and (c2) have masses with opposite signs 
 between the first and second generations,
 the mixing angle is stable 
 from the analogy of two-generation neutrinos.
 Therefore the maximal mixing between the first and second generations
 is conserved in the continuous region preserving the relation of 
 $\left| \phi _2 - \phi _1 \right| = \pi/2$.
As for stability of $\sin^2 2 \hat{\theta}_{13}$,
 Table \ref{tbl:C-MSWL} shows that the cases of (c3) and (c4) conserve 
 the zero mixing, while the cases of (c1) and (c2) do not.
Figure \ref{fig:C-MSWL}(b) shows that the stable region
 exists around the line of $\left| \phi _2 - \phi _1 \right| = 0$
 which connects (c3) and (c4), and
 the unstable region exists around the line 
 of $\left| \phi _2 - \phi _1 \right| = \pi/2$
 which connects (c1) and (c2).
For the stability of $\sin^2 2 \hat{\theta}_{23}$, 
 Table \ref{tbl:C-MSWL} shows that the case of (c3) only conserves
 the maximal mixing, and the case of (c4) induces the zero mixing.
Both cases of (c1) and (c2) induce $\sin^2 2 \hat{\theta}_{23} \sim 0.36$.
These situations are connected continuously by 
 two Majorana phases $\phi _1$ and $\phi _2$ 
 as shown Fig.\ref{fig:C-MSWL}(c).

\begin{figure}[tb]
  \begin{center}
    {\scalebox{0.8}{\includegraphics{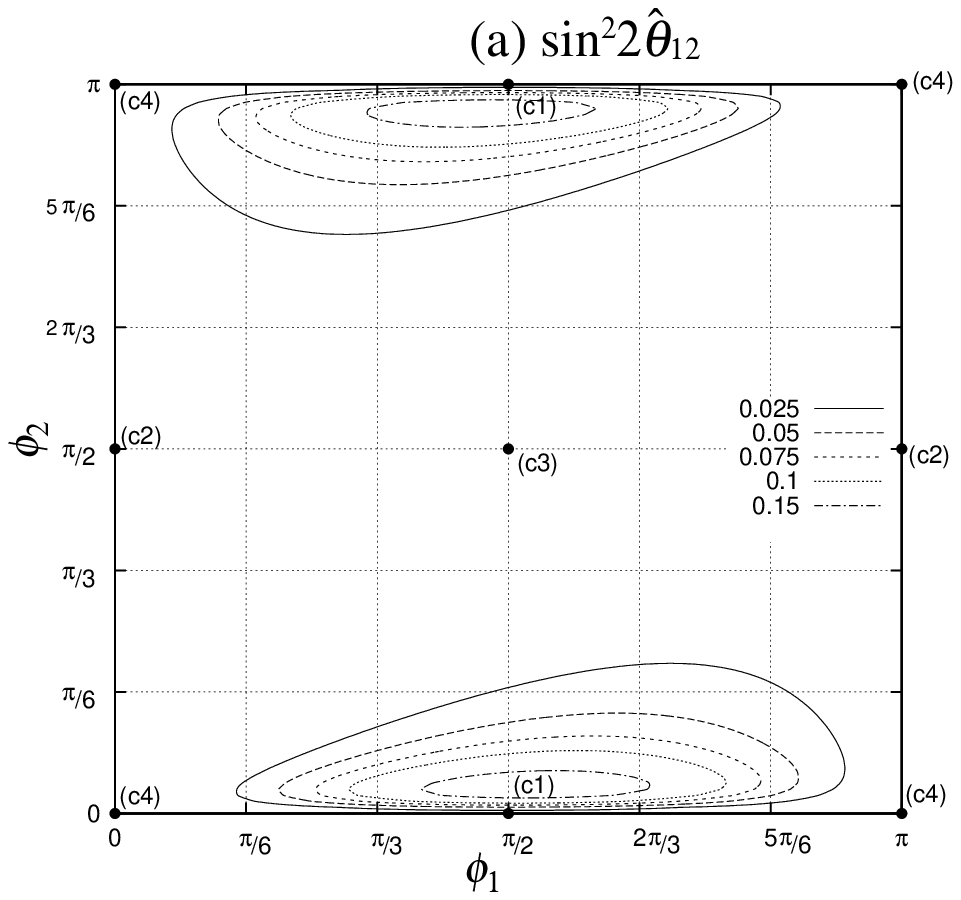}} } 
    {\scalebox{0.8}{\includegraphics{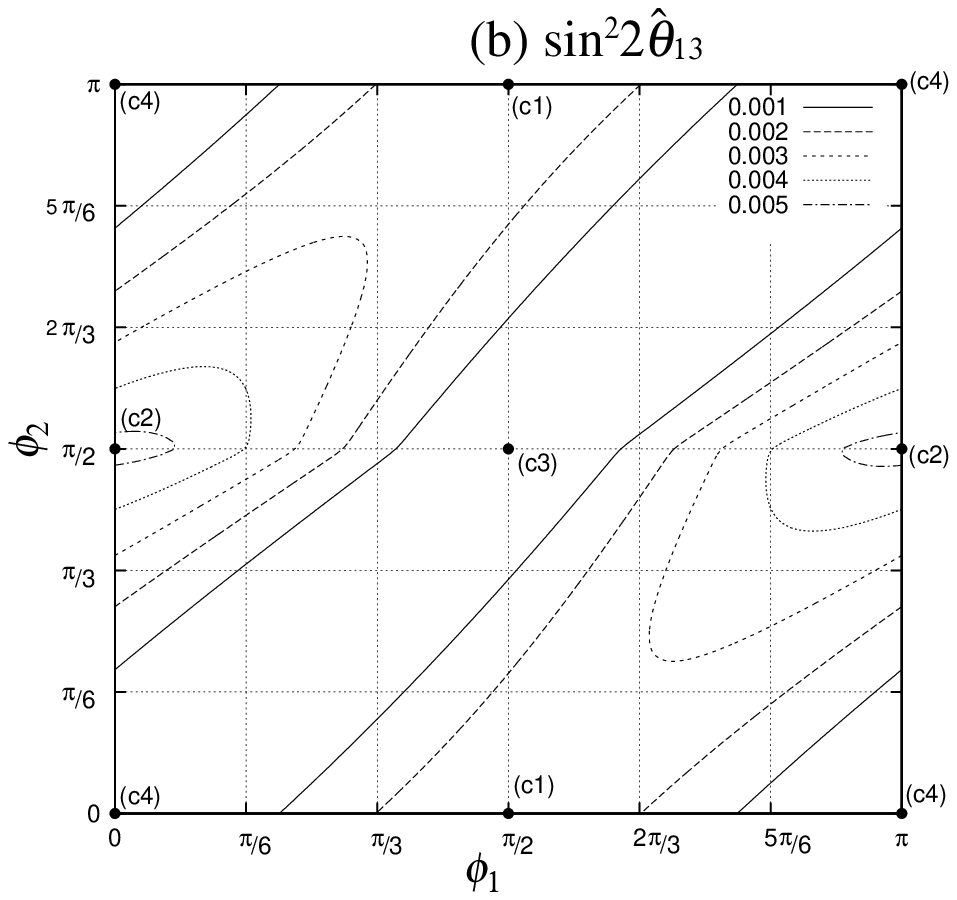}} } 
    {\scalebox{0.8}{\includegraphics{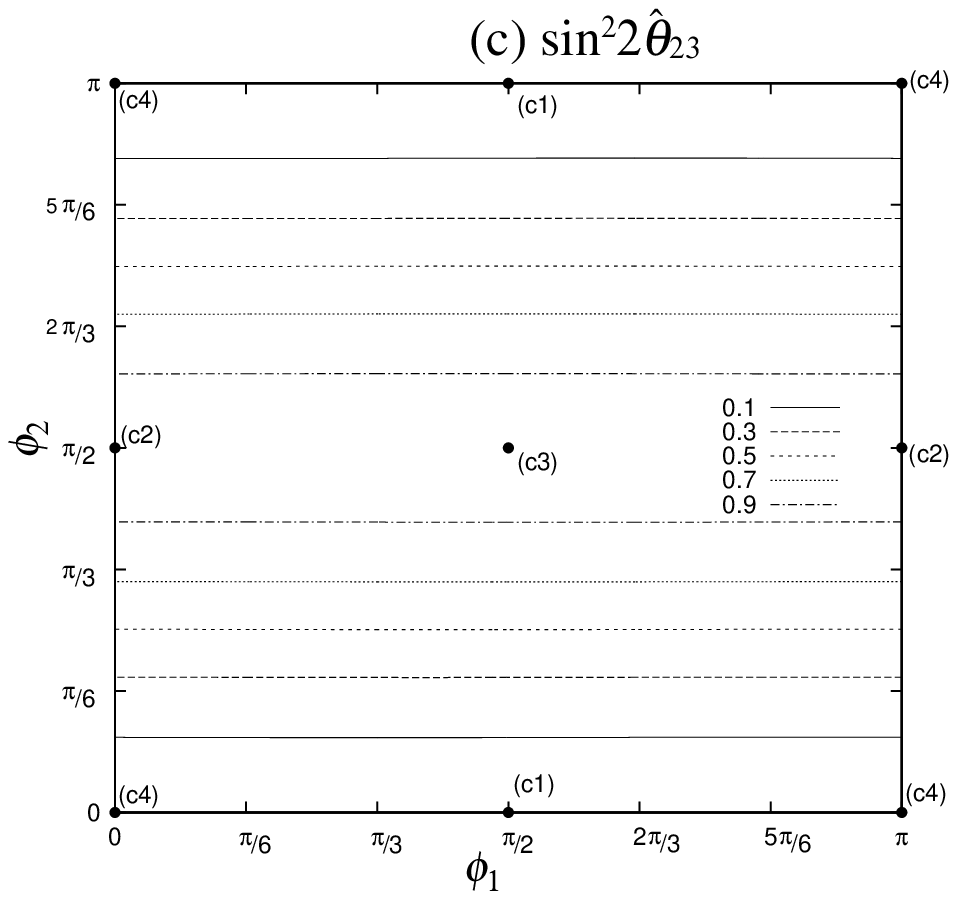}} } 
  \end{center}
  \vspace{-1.5em}
  \caption{The contour plots of (a): $\sin^2 2 \hat{\theta} _{12}$,
          (b): $\sin^2 2 \hat{\theta} _{13}$ and 
          (c): $\sin^2 2 \hat{\theta} _{23}$, 
          at $\mr = 10^{13}$ GeV in the case of
          the MSW-S solution with $\tan \beta = 60$. }
  \label{fig:C-MSWS} 
\end{figure}
Figures \ref{fig:C-MSWS} shows the values of mixing angles
 at high energy scale $\mr = 10^{13}$ GeV 
 for the continuous change of the Majorana phases 
 for the MSW-S solution in the case of $\tan \beta = 60$.
Table \ref{tbl:C-MSWS} shows the fixed values 
 of the mixing angles for the MSW-S solution
 in the large limit of quantum corrections,
 which are obtained from Eqns.(\ref{eqn:Cc1}) $\sim$ (\ref{eqn:Cc4}) by
 using Eqns.(\ref{sin2_12}) $\sim$ (\ref{sin2_23}).
%
\begin{table}[htb]
\begin{center}
\begin{tabular}{|c||c|c|c|c|}
\cline{2-5}
\multicolumn{1}{c|}{} & \multicolumn{1}{c|}{(c1)} & (c2) & (c3) & (c4) \\
\hline
$\sin^2 2 \hat{\theta}_{12}$ & $0.0$ & $0.0$ & $0.0$ & $0.0$ \\
\hline
$\sin^2 2 \hat{\theta}_{13}$ & $0.0$ & $0.0$ & $0.0$ & $0.0$ \\
\hline
$\sin^2 2 \hat{\theta}_{23}$ & $0.0$ & $1.0$ & $1.0$ & $0.0$ \\
\hline
\end{tabular}
\end{center}
\caption{The fixed values of the mixing angles for the MSW-S
         solution in the large limit of quantum corrections given by 
         Eqns.(\ref{eqn:Cc1}) $\sim$ (\ref{sin2_23}).
}
\label{tbl:C-MSWS}
\end{table}
The deviations from the values at $\mz$ scale, 
 $\sin^2 2 \theta _{12} = 7.1 \times 10^{-5}$, $\sin^2 2 \theta _{13} = 0$ and
 $\sin^2 2 \theta _{23} = 1$, indicate that the mixing angles receive
 significant quantum corrections.
For $\sin^2 2 \hat{\theta}_{12}$, Table \ref{tbl:C-MSWS} shows
 that all the cases of (c1) $\sim$ (c4) make it zero
 in the large limit of quantum corrections.
Figure \ref{fig:C-MSWS}(a) shows that the unstable region of
 $\sin^2 2 \hat{\theta}_{12} > 0.2$ exists around the points of 
 $(\phi _1 \;, \phi _2) \simeq (\pi /2, \pi / 30)$ and 
 $(\phi _1 \;, \phi _2) \simeq (\pi /2, 29 \pi / 30)$.
For the stability of $\sin^2 2 \hat{\theta}_{13}$,
 Table \ref{tbl:C-MSWS} shows that all 
 the cases of (c1) $\sim$ (c4) conserve the zero mixing.
Figure \ref{fig:C-MSWS}(b) shows that $\sin^2 2 \hat{\theta}_{13}$
 is stable with respect to
 the quantum corrections for any values of two Majorana phases.
For the stability of $\sin^2 2 \hat{\theta}_{23}$
 Table \ref{tbl:C-MSWS} shows that the cases of (c2) and (c3)
 conserve the maximal mixing, while the cases of (c1) and (c4)
 do not, in the large limit of quantum corrections.
Figure \ref{fig:C-MSWS}(c) shows that the stable region 
 exists around lines of $\phi _2 = \pi/2$
 which connect (c2) and (c3) by changing $\phi _1$ from $0$ to $\pi/2$, 
 and the unstable region exists
 around the lines $\phi _2 = 0$ and $\phi _2 = \pi$ which 
 connect (c1) and (c4) by changing $\phi _1$ from $0$ to $\pi/2$.

\section{Summary}

Neutrino-oscillation solutions for
 the atmospheric neutrino anomaly and the solar neutrino deficit
 can determine the texture of the neutrino mass matrix
 according to three types of neutrino mass hierarchies \cite{Altarelli} as
 Type A: $m_1^{} \ll m_2^{} \ll m_3^{}$,
 Type B: $m_1^{} \sim m_2^{} \gg m_3^{}$ , and
 Type C: $m_1^{} \sim m_2^{} \sim m_3^{}$.
We found that 
 the relative sign assignments of neutrino masses in each type of
 mass hierarchies play the crucial roles for the stability against
 quantum corrections. 
Actually, two physical Majorana phases in the lepton
 flavor mixing matrix connect among
 the relative sign assignments of neutrino masses. 
Therefore, in this paper we analyze 
 the stability of mixing angles against quantum corrections 
 according to three types of neutrino mass hierarchies
 (Type A, B, C) and  two Majorana phases. 
Two phases play the crucial roles for the stability of 
 the mixing angles against the quantum corrections.
The results in Ref.\cite{HO1}, where the stabilities of the 
 mixing angles in (a1) and (a2), (b1) and (b2), (c1) $\sim$ (c4)
 with respect to quantum corrections are argued,
 are reproduced by taking the definite values of
 two Majorana phases.

\section*{Acknowledgment}
The work of NO is supported by the JSPS Research Fellowship
for Young Scientists, No.2996.

%
%

\end{document}